\documentclass{aa}

\usepackage{hyperref}
\usepackage{graphicx}
\usepackage{txfonts}
\begin{document}

    \title{A semiempirical approach to low-energy cosmic ray propagation\\ in the diffuse interstellar medium}


    \author{Riccardo Franceschi \inst{1, 2} and 
        Steven N. Shore \inst{1,3}
    }

    \institute{Dipartimento di Fisica "Enrico Fermi", Universit\'a di Pisa , Largo B. Pontecorvo 3, 56127 Pisa, Italy
        \and Max-Planck-Institut f\"ur Astronomie (MPIA), K\"onigstuhl 17,  69117 Heidelberg, Germany 
        \and INFN - {2013} Sezione di Pisa, largo B.Pontecorvo 3, 56127 Pisa, Italy \\
    \email{franceschi@mpia.de,steven.neil.shore@unipi.it}
    }

    \date{}


    \abstract
    {We investigate the ionization of the diffuse interstellar medium by cosmic rays by modeling their propagation along the wandering magnetic fields using a Monte Carlo method. We explore how particle trapping and second-order Fermi processes affect the ionization of the medium.}
    {We study how low-energy cosmic rays propagate in turbulent, translucent molecular clouds, and how they regulate the ionization and both lose and gain energy from the medium.}
    {As a test case, we use high spatial resolution (0.03~pc) CO maps of a well-studied high latitude translucent cloud, MBM~3, to model turbulence.  The propagation problem is solved with a modified Monte Carlo procedure that includes trapping, energization, and ionization losses. }
    {In the homogeneous medium, trapping and re-energization do not produce a significant effect. In the nonuniform medium, particles can be trapped for a long time inside the cloud. This modifies the cosmic ray distribution due to stochastic acceleration at the highest energies ($\sim$ 100 MeV). At lower energies, the re-energization is too weak to produce an appreciable effect. The change in the energy distribution does not significantly affect the ionization losses, so ionization changes are due to trapping effects.}
    {Our Monte Carlo approach to cosmic ray propagation is an alternative method for solving the transport equation. This approach can be benchmarked to gas observations of molecular clouds. Using this approach, we demonstrate that stochastic Fermi acceleration and particle trapping occurs in inhomogeneous clouds, significantly enhancing their ionization.}

    \keywords{numerical methods  -- ISM: clouds -- ISM: cosmic rays -- turbulence}
\titlerunning{Cosmic ray propagation in the turbulent ISM}
    \maketitle
%

    \section{Introduction}
        The densest components of the Galactic interstellar medium (ISM), molecular clouds (MCs), have sufficiently neutral hydrogen high column densities, $N_H \sim 10^{18}-10^{21}$ cm$^{-2}$, which render them opaque  to ambient ionizing radiation. Thus,  hydrogen-ionizing photons are confined to the immediate surroundings of their stellar sources within clouds capable of star formation, or to a superficial, thin photodissociation and photoionized layer.  At greater depth in the clouds, cosmic rays (CRs) control much of the ionization, chemistry, and thermal balance~\citep{dalgarno06, padovani17}. The higher the ionization, the more the medium is coupled to the magnetic fields, which inhibits ambipolar diffusion and the formation of dense cores~\citep{ciolek93}. If, on the other hand, the ionization is low, the gas can drift across the field lines, allowing it to collapse faster~\citep{padovani09}. The thermal state of the gas is also affected by CR penetration, as the electrons freed though the charge-induced ionization transform their excess energy into heating and further ionize the gas.
        
        
        Cosmic rays  are thought to be accelerated by shock fronts, most likely supernova remnants (e.g., \citealt{Beresnyak09}), and then propagate in the ISM along magnetic field lines. They propagate gyrating around these field lines, but they can also be scattered by magnetic field inhomogeneities, local turbulence, or streaming instabilities generated by the CR propagation itself. Cosmic rays can be accelerated when scattered by the so-called Fermi acceleration mechanism. This mechanism was first proposed by \cite{Fermi49, Fermi54}, who speculated that CRs can be accelerated by random magnetic scatterers (or mirrors), typical of the turbulent ISM. The mirrors are large-scale fluctuations of the magnetic field ($\delta B / B \sim 1$) created by turbulent gas flows. If the magnetic mirror moves toward the particle, the particle will have increased energy after the reflection. By same account, the particle will lose energy if the magnetic mirror moves away from the particle. If the magnetic fluctuations are isotropic, the rate of a head-on collision is higher than a tail-on collision, and on average particles gain energy. This process is commonly known as stochastic (or second-order) Fermi acceleration. The effects of particle re-energization on the propagation of CRs have been analyzed in detail in the literature in the case of diffusion through shock fronts (e.g., \citealt{Perri12, Zimbardo15, Sioulas20}), but not so much for the diffuse ISM. However, ionization energy losses are strongly dependent on the energy of the traveling particle. By accelerating particles, stochastic Fermi acceleration could affect the ionization of the diffuse ISM by CRs.
        
        Understanding the effects of scattering on magnetic field inhomogeneities is essential to the description of CR propagation. At low densities, characteristic of diffuse MCs, the propagation of low-energy CRs is mostly determined by resonant scattering on magnetic fluctuation generated by the particles themselves, on the scale of the particle gyroradius. The challenge of studying CR propagation is its nonlinear nature, and while several studies have tackled this question (see, e.g., \citealt{Everett11, Morlino15, Ivlev18}), the solutions of the nonlinear system of equations describing the CR propagation are approximate  estimates  based on different assumptions. Different theoretical approaches have led to different, and at times opposite, results. For instance, \cite{Everett11} argue that the streaming instability (the Alfv-{00E9}n waves generated resonantly by the CRs themselves) excludes CRs from the interior of MCs. To the contrary, \cite{Padoan05} instead conclude that these instabilities enhance the CR penetration in the clouds.  The different methods used to solve the equation describing the CR propagation make it difficult to compare these results: \cite{Everett11}  solved  the hydrodynamical equation with spatial diffusion for the CRs pressure, while \cite{Padoan05} solved the kinetic transport equation.
        
        In this paper we explore the idea that the stochastic Fermi acceleration of charged particles affects the ionization state of the diffuse ISM. To test this hypothesis, it is not necessary to find the exact solution to the equation system for the CR propagation, which is beyond the scope of this paper. Instead, we present a new, semiempirical approach to the modeling of CR propagation based on CO observations of translucent MCs. We assume energy equipartition between the velocity field (derived from CO observations) and magnetic fields, and that the cloud has reached its equilibrium state with the incoming external radiation. This allows us to estimate the structure of the magnetic field inside the MC while automatically taking into account the nonlinear effect of the change in the magnetic fields caused by the passage of CRs. As a proof of concept, we study the simpler case of the propagation of low-energy protons ($E \lesssim 100$~MeV) in a turbulent diffuse cloud ($n \sim 100~ $cm$^{-3}$). To use a physically realistic structure for the turbulent field, we assume it to be the same as in the diffuse molecular cloud MBM~3. This is a well-studied source for which there are high spatial (0.03~pc) and high velocity (0.03~km/s) resolution CO observations, whose turbulent structure has been studied in detail, and which lacks evidence of internal sources or ongoing star formation processes \citep{shore06}.
       
        The reader should keep in mind that the purpose of this test is not an accurate modeling of the CR propagation inside MBM~3 (whose geometrical structure we do not try to reproduce). Rather, it is to use a MC with strong turbulence and no evidence of other ongoing physical processes as a benchmark to study CR propagation based on the turbulent gas structure. This approach is aimed at the problem of CR propagation in the ISM. When dealing with more complex structures, such as strong external radiation or internal star formation processes, the assumption of turbulence being in equilibrium with the magnetic field structure is no longer justified, and a proper solution of the  CR transport equation is needed to find the magnetic field structure. It is likely that the results would still be applicable for denser gas with a full treatment of the ionization balance, but this is unlikely in the case of star-forming clouds or clouds with a strong, ordered field. Nonetheless, including such effects from simulations could follow the same outline.

    \section{The model}
    \subsection{Physical bases of the model}
        The transport of charged particles is governed by their interaction with local electromagnetic fields, and the propagation of low-energy protons can be studied in the adiabatic limit, that is, when the particle gyroradius,  $r_g = mcv_\perp / qB$, where $v_\perp$ is the velocity component perpendicular to the magnetic lines of force, is much smaller than the characteristic scale over which the magnetic field changes, and the protons propagate along the magnetic lines of force. The effect of a change in the fields on a scale smaller than $r_g$ will be averaged. This condition is met for low-energy CRs propagating in diffuse clouds, where the typical magnetic field strength is a few $\mu G$~\citep{hennebelle12}. Two dominant effects  contribute to transport and diffusion: field line random walk (FLRW) and resonant wave-particle scattering. In the presence of FLRW alone, particles follow magnetic field lines and suffer spatial diffusion as the magnetic field lines diffuse in space~\citep{jokipii68}. This gives rise to an energy-independent diffusion of particle trajectories. In resonant wave-particle scattering, by the time a particle completes its orbit around the local magnetic field lines, it travels a distance along them approximately equal to the wavelength of an incoming wave. The force exerted by the magnetic perturbation on the particle, in this case, maintains the same direction over the interaction time, scattering the particle. The propagating particles can generate the waves that will scatter the particles themselves \citep{farmer04}. In this case, the propagation of charged particles is more complicated  since the process is nonlinear. Previous works~\citep{melrose74, drury83} showed how electrons are scattered by self-generated waves. However, protons do not interact with self-excited waves since they are much more massive. Their propagation can therefore be described using pure FLRW.  We will show how the effects of trapping and re-energization can be captured with a simple  semiempirical approach to the field line forcing through ambient cloud turbulence.  The nonlinear effects involved in electron propagation can be neglected.  By ``trapping'' we mean how inhomogeneities inside the medium  increase the particles' effective path. Re-energization is the exchange of energy between particles and macroscopic-scale turbulent flows. Any change in the magnetic field acts as a scattering center for the propagating particles. This causes an energy exchange due to the Fermi mechanism:

    \begin{equation}
        \Delta E/E \simeq 2 v_c v \cos \theta/c^2 \,,
    \end{equation}

    \noindent
    where $v$ is the particle velocity, $v_c$ is the scatterer velocity, and $\theta$ is their relative direction. A change in the CR energy distribution could affect the ionization state of the medium due to its dependence on the particle energy.

    \subsection{Turbulence in the diffuse medium}

    The transport of CRs inside MCs is driven by the stochastic nature of the internal magnetic fields, which leads to the diffusion of particles in both space and energy~\citep{thornbury14}. Turbulence is the principal agent for producing the complex structures in the ISM. A difficulty arises, however, in relating observation and turbulent theories because of the line-of-sight-integrated nature of the observations. This projection makes local quantities hard to derive, and the passage from the observational data to models is often based on assumptions. We assume the turbulence is homogeneous and in a state of equipartition between magnetic and kinetic turbulent energy.\par

    Intermittency at the dissipation scale is an essential characteristic of turbulence. This involves the occurrence of rare, large-amplitude events that are much more frequent than in a Gaussian process. Turbulent flows have been extensively studied in the laboratory, and in all cases the velocity distributions appear to be non-Gaussian. An extensive discussion on the observational signatures of interstellar turbulence can be found in~\cite{miesch99}. In this paper, we use these signatures to characterize interstellar turbulence. The use of observationally derived velocity probability distribution functions (PDFs) allows us to avoid a formal derivation of the turbulent spectral distribution.\par

    Our model is based on the study of the high latitude molecular cloud MBM~3 by~\cite{shore06}. In particular, we made use of the velocity centroid PDF obtained through CO observations. The centroid PDF is the distribution of the mean velocities of the line profiles taken over a large spatial sample. If $\Delta v(x,y) = v(x,y) - v_M(x,y)$ is the velocity fluctuation at any point $(x,y)$ and $v_M(x,y)$ is the mean velocity, then $\Delta v$ is the velocity centroid PDF.\par 


    Another useful PDF is the one constructed by the velocity difference for regions separated by a given spatial scale, or ``lag.'' This PDF is commonly used in the study of incompressible turbulence since the spatial lag determines the correlation between velocity flows in a Kolmogorov-like turbulent cascade. Classical turbulence theory requires that dissipation only occurs at the lower scale, where molecular viscosity becomes important. Below this scale, the fluctuations become uncorrelated, and we can safely assume that their effect on CR propagation averages to zero. By observing the lag at which the velocity centroid shift PDF relaxes to a Gaussian, we can evaluate the correlation length of the turbulent cascade. In our model we use as reference value the one measured in MBM~3 by~\cite{shore06}: $\ell_{corr} \sim 0.1$~pc.\par


    Assuming a standard Kolmogorov cascade, the turbulent energy transfer rate,

    \begin{equation}
        \epsilon = \rho  \langle \delta v^2 \rangle^{3/2} \ell^{-1}
        \label{eq:epsilon}
    ,\end{equation}

    \noindent
    is scale independent, where $\langle \delta v^2 \rangle$ is the velocity dispersion at the associated length, $\ell$. We can estimate $\sigma_v$ from the line profiles and take $\ell \approx 0.1$~pc. Given the characteristic number density of the diffuse medium $n \sim 100$ cm$^{-3}$, we have $\epsilon \sim 3 \times 10^{-23}$ erg. This quantity associates the observed velocity structures with the corresponding spatial structure of the turbulence. Assuming energy equipartition between the velocity and magnetic fields, Eq.\ref{eq:epsilon} can be used to derive the coherence structure of the magnetic lines of force from the velocity PDF.

    \subsection{Sampling the PDF}
    \label{sec:pdf}

    To reproduce the observational data, we took a medium with a characteristic dimension of 1~pc and number density 100~cm$^{-3}$. The turbulence coherence length ranges from 0.1 to 0.8~pc with an energy transfer rate $\epsilon =  3 \times 10^{-23}$. By changing the local physical conditions, we are able to explore their effect on the proton propagation. For instance, by reducing the mean free path in certain subregions of the medium, we are able to study the effects of particle trapping, and changing the turbulence PDF allows us to explore the effect of re-energization.\par

\begin{figure}[h]
        \centering
        \includegraphics[width=\hsize]{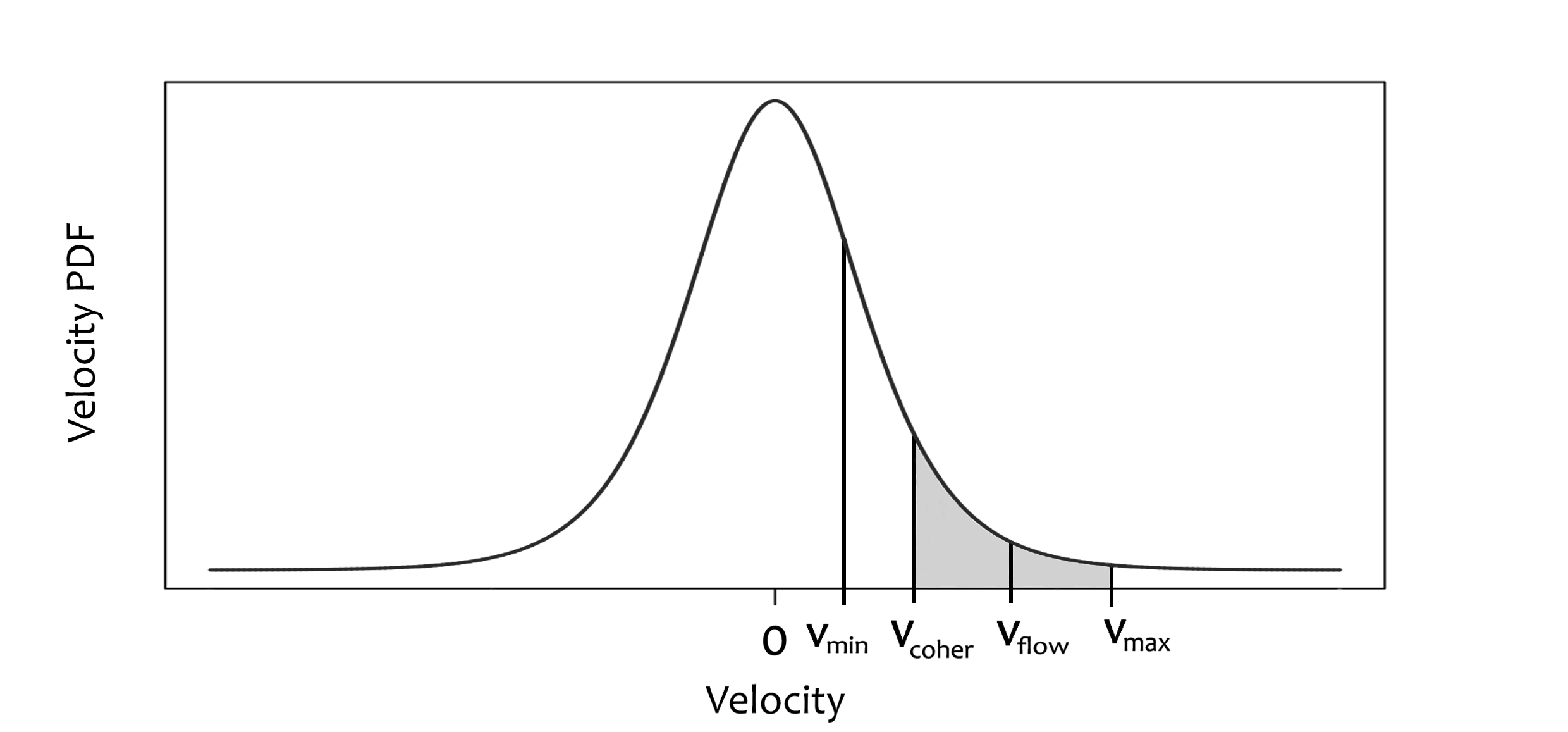}
        \caption{Visual representation of the method used to sample the length of each step and the velocity associated with the corresponding turbulent perturbation, as discussed in the text.}
        \label{Fig:pdf}
    \end{figure}

    The key point here is to use the velocity PDF to sample the distance a proton has to travel to meet a variation in the magnetic field. The adopted procedure is represented in Fig. \ref{Fig:pdf}. We started by sampling the velocity PDF to get the velocity of the scatterer, $v_{coehr}$, in between the turbulent velocity range (the gray region in the figure). This is the velocity of the fluid volume that carries the perturbation, which is necessary for computing the energy change due to the Fermi mechanism. We then used Eq. (\ref{eq:epsilon}) to find the coherence length of the perturbation. The proton has to travel a distance at least equal to this coherence length to change direction. However, it is unlikely for the particle to encounter the perturbation as soon as it travels this distance, as it can be caught by a larger-scale flow. Therefore, we sampled a coherence length again for the large-scale flow from the velocity PDF. This gives us the distance the proton crosses until the next change of direction.

    \section{Results}

    This approach can be used to simulate the trajectories of a large sample of protons. Along with the proton trajectories, we also computed the energy changes of traveling protons due to the Fermi mechanism and the ionization processes. We focused on the proton component of CRs as they compose about 90\% of CRs. The total ionization of the medium is also affected by electron propagation, mainly of secondary electrons produced by the first ionization events (e.g., \citealt{padovani09}), but here we are interested in how stochastic acceleration of CRs affects the ionization, rather than in modeling the total ionization of the medium.
    
    The data we used to compute the ionization losses can be found on the National Institute of Standards and Technology (NIST) website\footnote{https://physics.nist.gov/PhysRefData/Star/Text/PSTAR.html}. We began our analysis by simulating the trajectories of $10^4$ protons at $10$ MeV. The turbulence physical parameters are the ones derived from MBM~3 observations: the energy transfer rate is $\epsilon \sim 1.6 \cdot 10^{-12}$ km$^2$/s$^3$, and the cascade ranges from 0.1 pc to 0.8~pc, in a homogeneous medium of 1~pc size and number density $n \sim 100$~cm$^{-3}$. The corresponding turbulent velocity fluctuations are a few km/s (between 0.6 and 2.9~km/s).
    
\begin{figure}[h]
        \centering
        \includegraphics[width=\hsize]{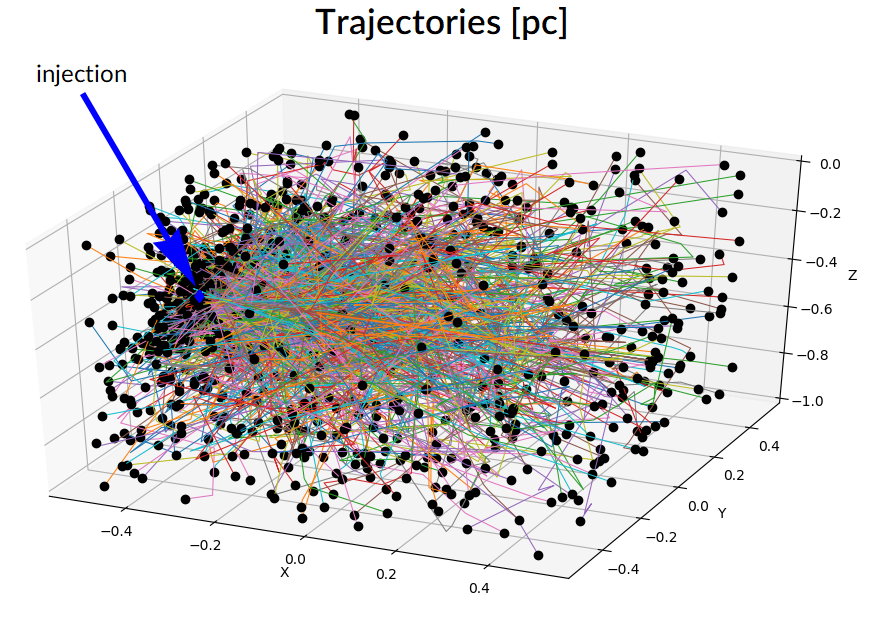}
        \caption{Trajectories of $10^3$ protons in a uniform medium. The blue circle is the injection point and black circles the escape points.}
        \label{fig:trajectory_uniform}
    \end{figure}

    In Fig. \ref{fig:trajectory_uniform} we show an example of the trajectories. The protons are all injected from the same point (in blue), and the  escape points are shown (black circles). The escaping points uniformly diffuse away from the injection point, as expected in a random walk.\par

\begin{figure}[h]
        \centering
        \includegraphics[width=\hsize]{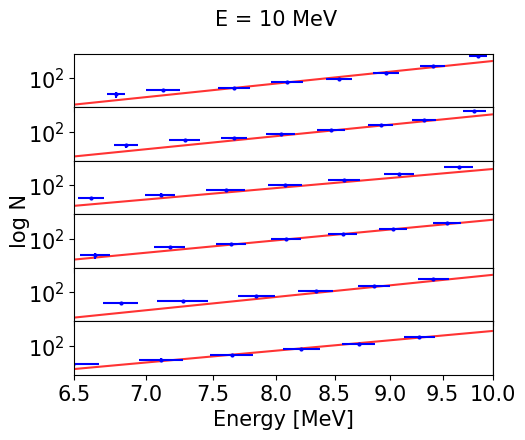}
        \caption{Histogram of 10 MeV protons propagating in a uniform medium. The top plot represents the particles on the injection face reflected by the medium. The middle plots are at increasing depth, and the bottom represents transmission. The red lines show a power-law fit of the particle distributions.}
        \label{fig:10mev_homo}
    \end{figure}

    An example of the energy distribution of the protons at different depths in the medium is shown in Fig. \ref{fig:10mev_homo}. Each point represents the mean energy of the protons within the energy bin. We used the variance of these energies as the error in energy, and for the error on the number of protons in a given bin we used a Poisson statistic $\sigma = \sqrt{N}$, where N is the number of protons inside the bin.
    
    Starting from a fixed initial energy of 10~MeV, we find that the particle energy distributions at different penetration depths in the medium are well fitted by the same power-law distribution (Fig. \ref{fig:10mev_homo}). This is an important consistency check as the  PDF -- a Cauchy distribution -- is characterized by a power-law tail. Moreover, the energy distribution is well represented by the same power law at every penetration depth in the medium, as one would expect since the PDF does not change within the medium. In Table \ref{tab:10_homo} we also give the proton ionization losses.


    \begin{table}[h!]
        \begin{center}
            \begin{tabular}{| c | | c c c c c |}
                \multicolumn{6}{c}{Ionization energy loss at 10 MeV [MeV]} \\
                \hline
                $1^{st}$ layer & 99.9  & 259.0 & 1547.2 & 262.6 & 103.0 \\
                \hline
                $2^{nd}$ layer & 125.6 & 284.1 & 980.8  & 288.4 & 124.5 \\
                \hline
                $3^{rd}$ layer & 102.4 & 226.5 & 531.6  & 229.4 & 109.4 \\
                \hline
                $4^{th}$ layer & 78.1  & 134.3 & 271.3  & 135.6 & 75.7  \\
                \hline
                $5^{th}$ layer & 41.8  & 65.8  & 122.3  & 65.0  & 32.2  \\
                \hline
            \end{tabular}
            \caption{Total ionization losses for $10^4$ protons at 10 MeV penetrating the medium. Each layer is a different depth within  the medium, the first being the injection face, the last the  escape face. Each layer is divided into five regions to get a spatial distribution for the ionization.}
            \label{tab:10_homo}
        \end{center}
    \end{table}

\begin{figure}[h]
        \centering
        \includegraphics[width=\hsize]{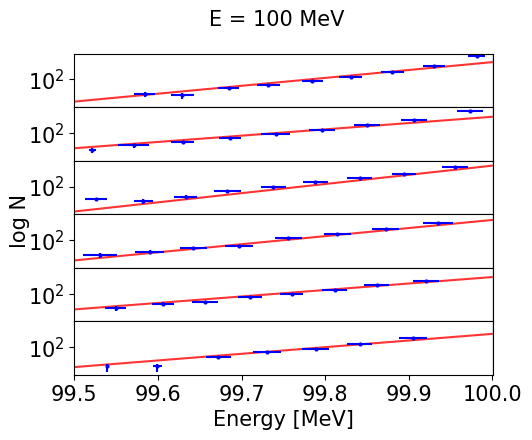}
        \caption{Histogram of 100 MeV protons propagating in a uniform medium. The top plot represents the particles on the injection face reflected by the medium. The middle plots are at increasing depth, and the bottom represents transmission. The red lines show a power-law fit of the particle distributions.}
        \label{fig:100mev_homo}
    \end{figure}

    At 100~MeV, we do not see any significant change in the results, as shown in Fig. \ref{fig:100mev_homo}. The main difference is that the energy spectrum gets harder, since the ionization losses are higher at lower energies and at the initial energy the final energy distribution is less extended. The resulting power-law tail is due to the lossy random walk and is an intrinsic result of our model. In Table \ref{tab:100_homo} we list the ionization losses at 100 MeV.

    \begin{table}[h!]
        \begin{center}
            \begin{tabular}{| c | | c c c c c |}
                \multicolumn{6}{c}{Ionization energy loss at 100 MeV [MeV]} \\
                \hline
                $1^{st}$ layer & 14.6 & 35.6 & 227.1 & 38.1 & 15.3 \\
                \hline
                $2^{nd}$ layer & 18.2 & 40.1 & 142.6 & 40.5 & 17.8 \\
                \hline
                $3^{rd}$ layer & 15.7 & 30.9 & 77.4  & 32.1 & 16.0 \\
                \hline
                $4^{th}$ layer & 9.6  & 18.9 & 39.6  & 18.8 & 11.2 \\
                \hline
                $5^{th}$ layer & 4.6  & 8.6  & 15.2  & 9.0  & 6.1  \\
                \hline
            \end{tabular}
            \caption{Ionization losses for $10^4$ protons at 100 MeV penetrating the medium. Each layer is at a different depth inside the medium, as in Table 1.}
            \label{tab:100_homo}
        \end{center}
    \end{table}

    \begin{figure}[h]
        \centering
        \includegraphics[width=\hsize]{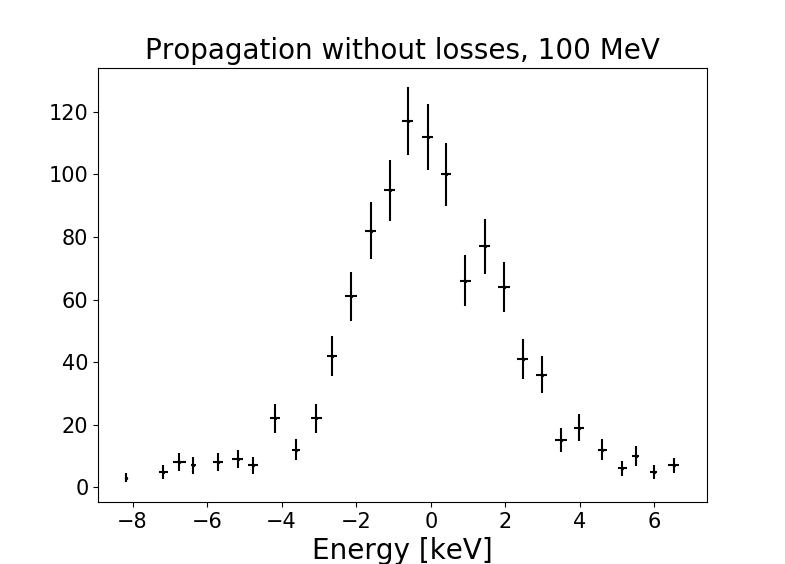}
        \caption{Proton energy distribution neglecting losses at half depth in the medium.}
        \label{fig:100_reen}
    \end{figure}

    These results are unaffected by re-energization to within the statistical uncertainties. To see its contribution, we removed the ionization losses and examined the particle energy distribution at half depth in the medium, as shown in Fig. \ref{fig:100_reen}. The re-energization has just a minor effect on the particle energy, around 1 keV for protons at 100 MeV energy and even less at lower energies. This seems to indicate that re-energization is negligible, but we have only considered a single low column density uniform medium. The ISM is, however, unlikely to be homogeneous, so in the following section we examine the effects of an inhomogeneous medium on the propagation of protons.

    \subsection{Effects of trapping}
\begin{figure}[h]
        \centering
        \includegraphics[width=\hsize]{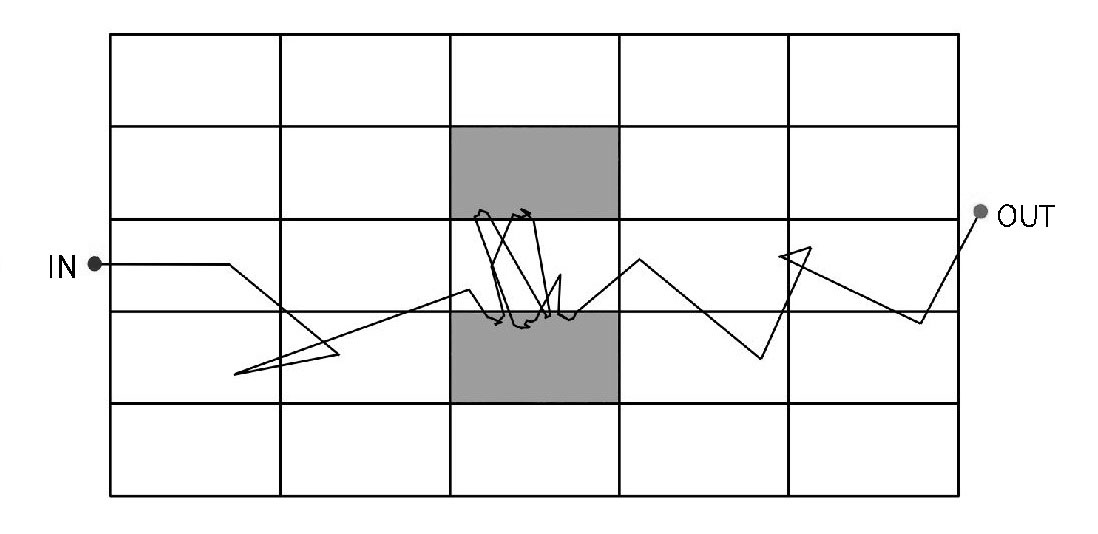}
        \caption{Configuration of the medium used to investigate the effects of inhomogeneities, represented by the gray regions.}
        \label{fig:dishomo}
    \end{figure}

    To study the propagation in a nonuniform medium, we added two denser regions, each with a proton mean free path much shorter than in the rest of the medium, with $\epsilon \sim 1.6 \cdot 10^{-6}$ km$^2$/s$^3$ and a turbulent cascade ranging from 0.01 to 0.1 pc. This implies flows up to 50 km/s, which enhance the amount of energy exchange between the flows and the protons. Although such enormous fluctuations are too large for individual clouds, they can be observed in structures of the diffuse medium: this is a schematic picture of a set of filaments in large-scale shear flows, such as the Herschel filaments~\citep{arzoumanian17}. The geometry adopted is shown in Fig. \ref{fig:dishomo}. In this case, a proton can be trapped between regions and scatter many times before eventually escaping. By increasing the number of scatterings, it is possible for the proton to undergo a large number of energizing events.
    
    
\begin{figure}[h]
        \centering
        \includegraphics[width=\hsize]{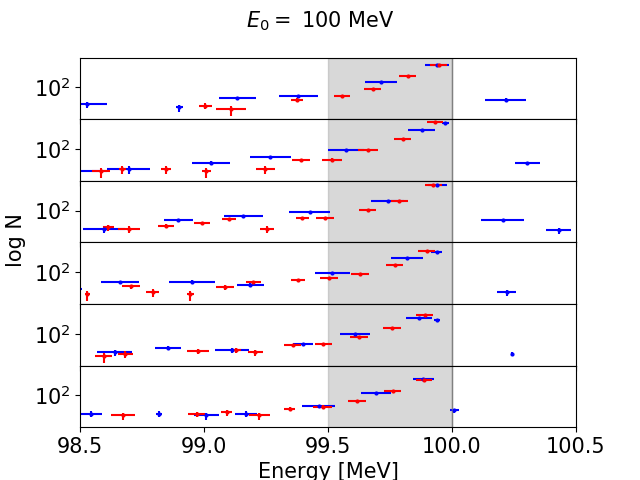}
        \caption{Histogram of 100 MeV protons propagating in an inhomogeneous medium. The top plot represents the particles on the injection face reflected by the medium. The middle plots are at increasing depth, and the bottom represents transmission. The third and fourth plots from the top represent the interfaces with the inhomogeneous region. For comparison, the gray area shows the particle energy range in a homogeneous medium. The blue and red points do and do not include the effects of re-energization, respectively. The flat distribution in the lower energy range is the signature of trapping-enhanced ionization losses and small re-energization.}
        \label{fig:100_trapping}
    \end{figure}

    The exiting energy distribution is shown in Fig.\ref{fig:100_trapping} for $10^4$ particles at 100~MeV, where the third and fourth plots from the top represent the proton energy distribution at the interfaces of the inhomogeneous region. We show the distributions that include the effects of re-energization (in blue) as well as the distribution without it (in red).

    \begin{table}[h]
        \begin{center}
            \begin{tabular}{| c | | c c c c c |}
                \multicolumn{6}{c}{Ionization energy loss at 100 MeV [MeV]} \\
                \hline
                $1^{st}$ layer & 9.0 & 19.8 & 157.5 & 19.6 & 9.4 \\
                \hline
                $2^{nd}$ layer & 9.9 & 25.3 & 107.9 & 26.4 & 9.8 \\
                \hline
                $3^{rd}$ layer & 2.9 & 29.5 & 100.1 & 34.1 & 3.5 \\
                \hline
                $4^{th}$ layer & 4.3 & 11.6 & 52.8  & 12.6 & 4.4 \\
                \hline
                $5^{th}$ layer & 3.8 & 8.1  & 28.7  & 7.9  & 3.5 \\
                \hline
            \end{tabular}
            \caption{Ionization losses for $10^4$ protons at 100 MeV penetrating a nonuniform medium. The inhomogeneities are placed in the third layer, according to Fig. \ref{fig:dishomo}}
            \label{tab:100_trapping}
        \end{center}
    \end{table}

    In the case of re-energization we see particles at higher energies than injection. The two inhomogeneous regions lead to the enhanced ionization of the medium relative to the uniform case (Table \ref{tab:100_homo}), seen in the third line of Table \ref{tab:100_trapping}. This implies that the incident CR flux is overestimated from the ionization state of the medium. Our model also shows that the presence of inhomogeneities is coupled with an excess of higher-energy CRs. However, the ionization is further affected by this change in the CR spectrum since it is energy dependent. The ionization of the medium is similar in the case of the homogeneous and inhomogeneous medium, with the main difference being on the line of sight passing between the inhomogeneities\par
    
    \begin{figure}[h]
        \centering
        \includegraphics[width=\hsize]{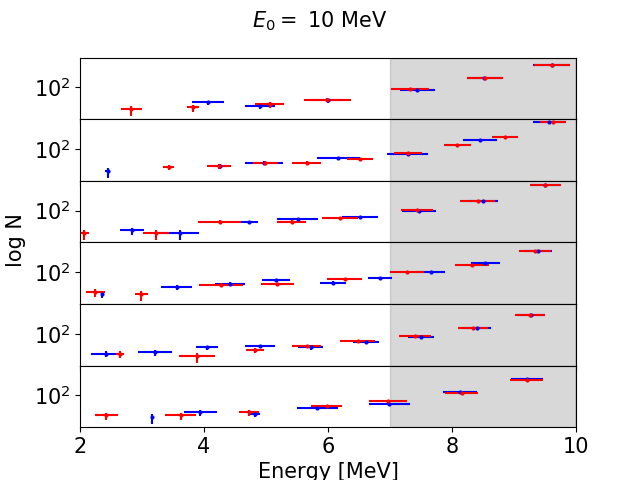}
        \caption{As in Fig. \ref{fig:100_trapping}, histogram of 10 MeV protons propagating in an inhomogeneous medium. }
        \label{fig:10_trapping}
    \end{figure}
    
    \begin{table}[h]
        \begin{center}
            \begin{tabular}{| c | | c c c c c |}
                \multicolumn{6}{c}{Ionization energy loss at 10 MeV [MeV]} \\
                \hline
                $1^{st}$ layer & 101.8 & 260.1 & 1504.2 & 262.0 & 102.9 \\
                \hline
                $2^{nd}$ layer & 117.6 & 328.0 & 1139.3 & 357.0 & 119.4 \\
                \hline
                $3^{rd}$ layer & 40.6  & 438.5 & 903.8  & 393.6 & 45.3  \\
                \hline
                $4^{th}$ layer & 45.2  & 128.2 & 426.3  & 127.3 & 49.8  \\
                \hline
                $5^{th}$ layer & 35.9  & 78.3  & 187.1  & 71.4  & 39.0  \\
                \hline
            \end{tabular}
            \caption{Ionization losses for $10^4$ protons at 10 MeV penetrating a nonuniform medium. The inhomogeneities are placed in the third layer, according to Fig. \ref{fig:dishomo}}
            \label{tab:10_trapping}
        \end{center}
    \end{table}

    In Fig. \ref{fig:10_trapping} we show that at 10~MeV the efficiency of re-energization is not enough to affect the energy distribution of the protons. Therefore, the main effect of trapping is to enhance the energy losses by increasing the effective path of protons inside the medium, in agreement with, for example, \cite{padovani09} and \cite{Morlino15}. The ionization losses, reported in Table \ref{tab:10_trapping}, are unaffected by re-energization effects.\par

    \subsection{Number of ionization events}
    A modified version of this algorithm was implemented by converting the ionization energy losses to the number of ionization events, neglecting secondary ionization by e$^-$ collision. As a proton exchanges energy with the gas, each ionization event will correspond to an energy loss equal to $I(H) = 13.598$~eV, the hydrogen ionization energy plus the kinetic energy of the ejected electron. The typical energy loss per ionization event depends on the proton energy. In the case of CR re-energization, a change in energy alters the number of ionization events even when the total energy loss does not change, as our simulation suggests. However, the energy of the ejected electron can extend just a few eV above the threshold and is a very minor effect. To estimate the average energy loss per ionization event, we used the cross section $\sigma_p^{ion}$ for hydrogen ionization by proton impact. The available experimental data have been summarized by \citet{rudd79}. The data were fitted with expressions appropriate to the high-energy and low-energy limit:
    
    \begin{equation}
        \sigma_p^{ion} = (\sigma_{low}^{-1} + \sigma_{high}^{-1})^{-1},
    \end{equation}where
    
    \begin{equation}
        \sigma_{low} = 4 \pi a_0^2 C x^D; \qquad \sigma_{high} = 4 \pi a_0^2 [ A \log(1+x) + B] x^{-1},
    \end{equation}with $a_0 =  5.29 \cdot 10^{-9}$~cm, $x = m_e E_p / m_p I(H)$, $I(H) = 13.598 \; eV$, $A = 0.71$, $B = 1.63$, $C = 0.51$, and $D = 1.24$. Using this expression, we find that re-energization does not change the number of ionization events in the medium at 10 MeV nor at 100 MeV.

    \subsection{Dense cores}
    In dense MCs, such as cold cores or dark clouds, the physical properties of the medium are governed by charged particles. By driving the ionization of the medium, they determine the gas coupling with the magnetic field. If the ionization is low, the gas is mostly neutral and free to stream through the field lines, favoring the collapse of the cloud and possibly triggering star formation processes \citep{padovani14}. We increased the number density of the medium to $10^4 \; cm^{-3}$ and used a particle mean free path one order of magnitude smaller than in the uniform case already shown  ($\approx 10^{-3}$~pc). The other parameters were left unchanged. We note that in dark clouds our model does not correctly describe the physical picture because it is not possible to derive the velocity PDF using the approach shown in Sect. \ref{sec:pdf}. Those clouds are often self-gravitating, and thus our description of turbulence is not appropriate. In this case we should take the effect of ambipolar diffusion, the decoupling of neutrals from the ionized gas, into account. This, with the lower ionization rate, allows neutrals to aggregate under the effect of self-gravity~\citep{ciolek93}. Moreover, stronger losses mean that higher-energy protons, which initially are not coupled to the turbulent magnetic field, slow  until their energy falls into the range of FLRW propagation~\citep{indriolo09, padovani09}. Our model only treats protons that remain bound to the field lines. However, the form of the turbulent PDF does not differ observationally, at least qualitatively, from denser translucent clouds. So, as a schematic model, we used our approach to simulate a dense core to explore density effects.\par
    
    \begin{figure}[h]
        \centering
        \includegraphics[width=\hsize]{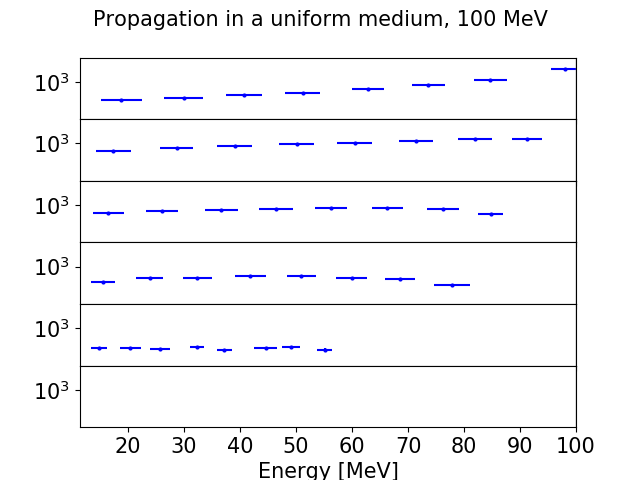}
        \caption{Histogram of 100~MeV protons propagating in a dense core.}
        \label{fig:core}
    \end{figure}

    \begin{table}[h]
        \begin{center}
            \begin{tabular}{| c | | c c c c c |}
                \multicolumn{6}{c}{Ionization energy loss at 100 MeV [GeV]} \\
                \hline
                $1^{st}$ layer & 2.3  & 10.3 & 45.5 & 9.9  & 2.1  \\
                \hline
                $2^{nd}$ layer & 2.6  & 8.6  & 15.1 & 7.8  & 2.4  \\
                \hline
                $3^{rd}$ layer & 1.0  & 2.7  & 4.9  & 2.6  & 1.0  \\
                \hline
                $4^{th}$ layer & 0.3  & 0.5  & 0.9  & 0.6  & 0.2  \\
                \hline
                $5^{th}$ layer & 0.06 & 0.10 & 0.13 & 0.06 & 0.06 \\
                \hline
            \end{tabular}
            \caption{Ionization losses for $10^4$ protons at 100 MeV penetrating a dense core.}
            \label{tab:core}
        \end{center}
    \end{table}

\begin{figure}[h]
        \centering
        \includegraphics[width=\hsize]{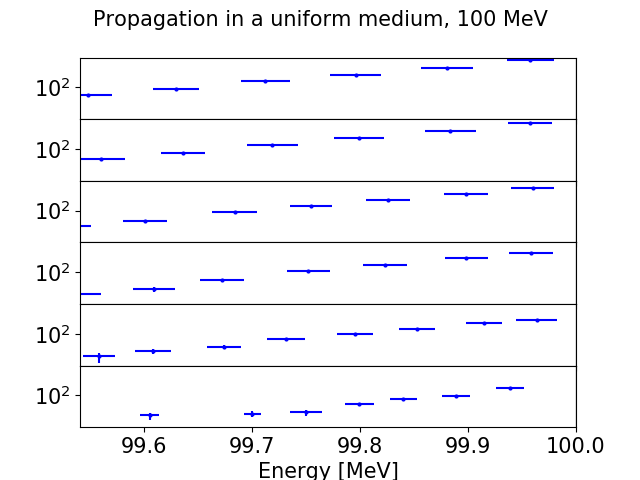}
        \caption{Histogram of 100 MeV protons propagating in a uniform medium with number density $300 \; cm^{-3}$.}
        \label{fig:3density}
    \end{figure}

    In Fig. \ref{fig:core} we show that much less energy is lost to the medium than in the diffuse case, as the propagation is hindered by the lower mean free path. Over a given path length, more energy is lost through ionization than in a more diffuse medium. However, few particles penetrate the medium, and they are more likely to be quickly reflected by it. As a result, they spend little time inside a dense medium, and the net ionization energy loss is low. For comparison, we also show the results for a less extreme case, using a number density of $300 \; cm^{-3}$. Figure \ref{fig:3density} and Table \ref{tab:3density} do not show results qualitatively different from the result for the diffuse medium, other than emphasized losses.
    
    \begin{table}[h]
        \begin{center}
            \begin{tabular}{| c | | c c c c c |}
                \multicolumn{6}{c}{Ionization energy loss at 100 MeV [MeV]} \\
                \hline
                $1^{st}$ & 33.7 & 161.5 & 723.7 & 166.8 & 36.2 \\
                \hline
                $2^{nd}$ & 43.4 & 174.8 & 427.5 & 181.3 & 45.7 \\
                \hline
                $3^{rd}$ & 36.3 & 117.1 & 213.5 & 116.5 & 38.8 \\
                \hline
                $4^{th}$ & 24.6 & 67.1  & 98.3  & 66.1  & 22.3 \\
                \hline
                $5^{th}$ & 11.7 & 26.3  & 37.5  & 26.5  & 11.1 \\
                \hline
            \end{tabular}
            \caption{Ionization losses for $10^4$ protons at 100 MeV in a medium with number density $300 \; cm^{-3}$.}
            \label{tab:3density}
        \end{center}
    \end{table}

    \subsection{Imbedded slab with two-sided incidence}
\begin{figure}[h]
        \centering
        \includegraphics[width=\hsize]{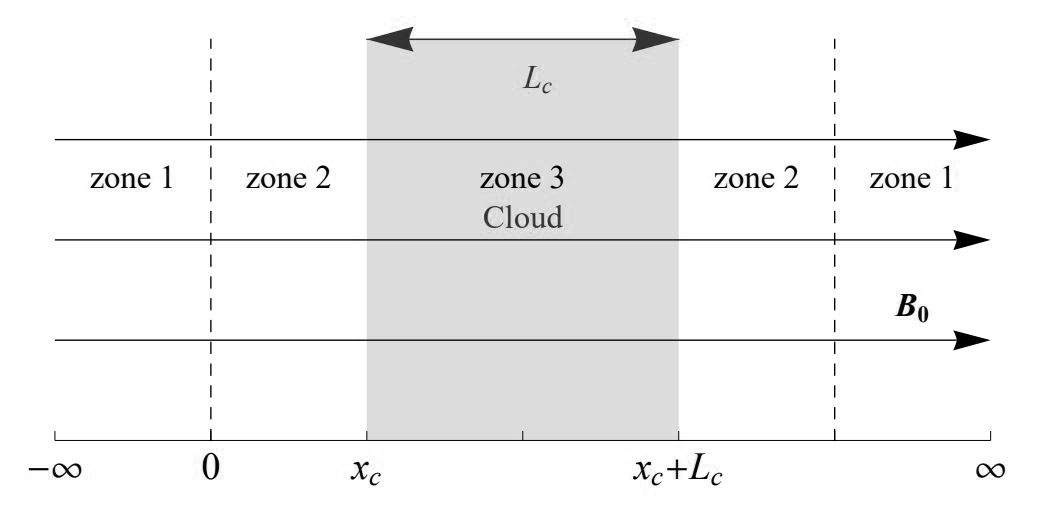}~\caption{Sketch of the simplified model used to describe the cloud geometry \protect{\citep{Morlino15}}.}
        \label{fig:sketch_morlino}
    \end{figure}

~\cite{Morlino15} studied the propagation of CRs in a dense core by dividing the ISM into three regions, shown in our Fig. \ref{fig:sketch_morlino}: (1) a zone far  from the cloud ($x<0$) where CRs are unaffected by the presence of the cloud, (2) a region immediately outside of the cloud ($0 < x < x_c$), and (3) a dense cloud   ($x_c < x < x_c + L_c$).
    
\begin{figure}[h]
        \centering
        \includegraphics[width=\hsize]{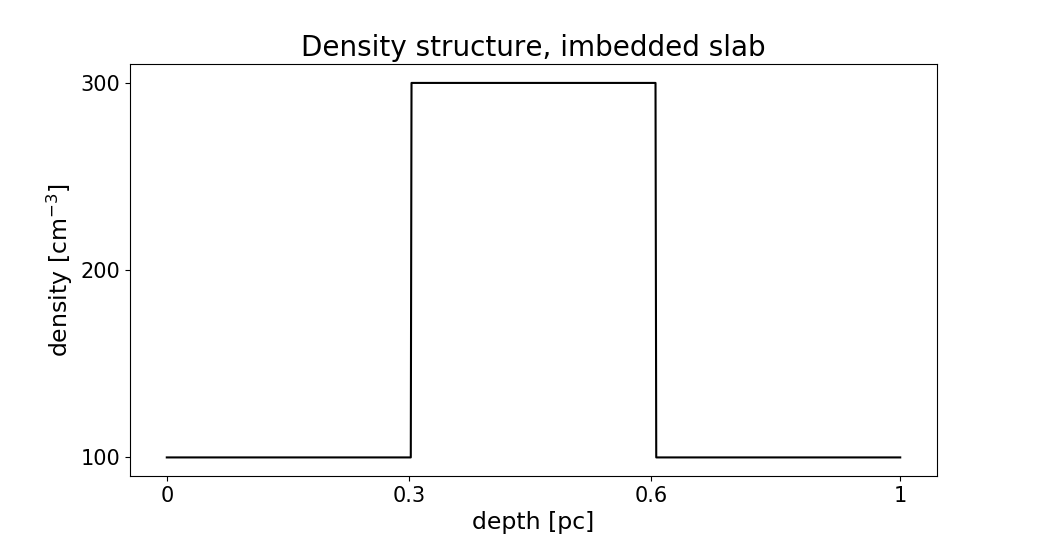}
        \caption{Density profile for the imbedded slab.}
        \label{fig:density_structure}
    \end{figure}

     They formally solved the transport equation. Here we show  the results of our alternative procedure. For zone (2) we used the parameter as the simulation of the uniform, diffuse medium, while for zone (3) the values are those of the dense core. The protons are injected from $-\infty$ and $+\infty$ (Fig. \ref{fig:density_structure}).\par
    
    \begin{figure}[h]
        \centering
        \includegraphics[width=\hsize]{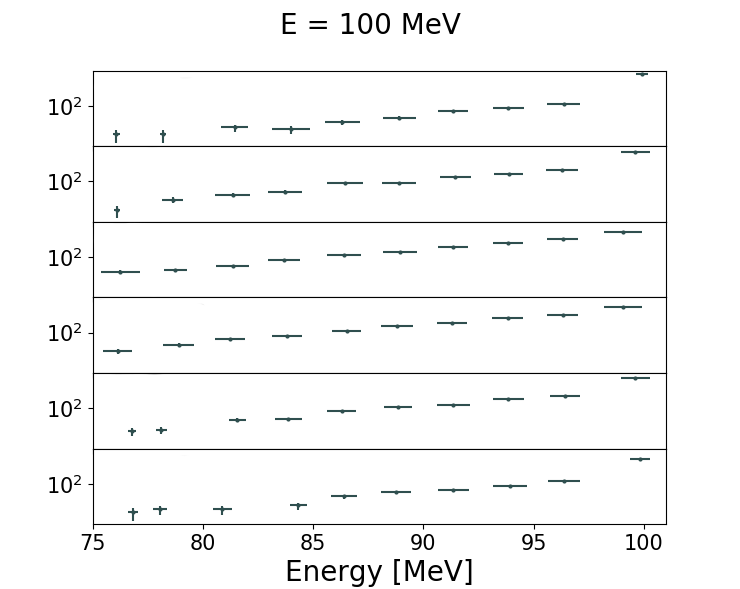}
        \caption{Histogram of 100 MeV protons propagating in a stratified medium:~\cite{Morlino15} model. The top two and bottom two  plots refer to the diffuse region, $n = 100 \; cm^{-3}$, and the third and fourth plots are for the dense embedded region, $n = 300 \; cm^{-3}$. Note that in the top and bottom panel, which show the distribution in the external medium, the low-energy flat spectrum is from protons that escape after partial trapping in the denser stratum. See also Figs. \ref{fig:100_trapping} and \ref{fig:10_trapping}.}
        \label{fig:morlino}
    \end{figure}
    
    \begin{table}[h]
        \begin{center}
            \begin{tabular}{| c | | c c c c c |}
                \multicolumn{6}{c}{Ionization energy loss at 100 MeV [MeV]} \\
                \hline
                $1^{st}$ & 26.0   & 93.5    & 384.7   & 93.5    & 22.8   \\
                \hline
                $2^{nd}$ & 35.8   & 127.1   & 265.7   & 131.4   & 41.9   \\
                \hline
                $3^{rd}$ & 4641.7 & 14946.1 & 24546.8 & 14964.9 & 4552.7 \\
                \hline
                $4^{th}$ & 39.2   & 137.7   & 278.9   & 131.1   & 36.3   \\
                \hline
                $5^{th}$ & 24.6   & 103.5   & 390.8   & 96.4    & 22.1   \\
                \hline
            \end{tabular}
            \caption{Ionization losses for $10^4$ protons at 100 MeV in a stratified medium.}
            \label{tab:morlino}
        \end{center}
    \end{table}
    
    \cite{Morlino15} find a reduction in the streaming velocity of the CRs to the Alfv\'en speed, $v_A$. This is a consequence of the {streaming instability}, the excitation of magnetohydrodynamic waves by particles moving faster than $v_A$. If charged particles in a cloud stream faster than the local Alfv\'en speed, they feed the Alfv\'en waves in the medium, which grow in amplitude and generate turbulence~\citep{amato18, bell04}. \cite{Morlino15} point out that the reduction in the CR flux in the cloud does not depend on the presence of a streaming instability, but is rather a consequence of the cloud structure. Although the solution of the problem requires a much broader range of processes, our point is that a  reduction in the CR flux can be further enhanced by this turbulent interaction. In addition, comparing Figs. \ref{fig:100mev_homo} and \ref{fig:morlino} shows that the flux of CRs in the diffuse region (2) is suppressed relative to a uniform diffuse medium. The presence of a dense intermediate stratum  enhances the energy losses and broadens the CR energy distribution, reducing their flux per unit energy, whereupon this reduced flux is fed back to the nearby diffuse region and reduces the CR flux in this region even without the generation of  magnetohydrodynamic waves (as in \cite{Morlino15}).

    \section{Conclusions}

    We have presented a methodology to explore the effects of trapping and re-energization on the interaction of low-energy CRs with a diffuse cloudy ISM. We explored these effects in the context of a physically justifiable, simplified model problem. We considered only the propagation of protons since the electrons' propagation is nonlinear; their role has to be assessed in a further work.\par
    An effect of re-energization is the change in the CR energy distribution. Since ionization losses depend on the particle energy, it can affect the total energy loss. In the dense medium case, where CRs can lose all their energy, re-energization changes the total amount of energy that can be lost via ionization. Our results suggest that the re-energization does not change the spectrum or the ionization state in a diffuse, uniform cloud. However, with inhomogeneities, the re-energization effects are enhanced, as particles trapped between inhomogeneities undergo a large number of interactions with the turbulent structure. For $\sim 100$ MeV protons, this results in an energy distribution that becomes more extended toward both  lower and higher energies. At $\sim 10$ MeV, proton re-energization does not affect the distribution, since the ionization losses are higher and dominate its effects.\par
    The CR trapping raises the energy loss by increasing the effective column density traversed. This result is due to the geometry of the medium, so it affects protons at both 10 MeV and 100 MeV. In contrast, the change in the CR spectrum  at 100 MeV is too small to affect the loss term.\par
    As a final test we examined the configuration studied by~\cite{Morlino15}, which consists of a MC surrounded by a more diffuse medium. Although in their study they accounted for nonlinear effects, we were able to qualitatively reproduce their CR energy distribution. In particular, we find that the flux of CRs is suppressed in the proximity of the cloud. The presence of the MC enhances the energy losses, resulting in a reduced flux of CRs per unit energy. This effect is caused by the assumed geometry and does not require nonlinear feedback effects, but the excitation of Alfv\'en waves will further prevent CRs from penetrating the cloud and should be included in a more extended treatment. \par

    \begin{acknowledgements}
        Special thanks go to Daniele Galli and Marco Padovani for their useful insight on the general direction of this work.  A portion of this work was performed while SNS held a visiting professor at the Astronomical Institute of Charles University.  We thank David Vokrouhlick-{00FD}, the department chair, for the invitation and support for a study period for RF.  
        RF also thanks his MSc thesis committee members, Luca Baldini and Massimiliano Razzano, for their their feedback on this project.
    \end{acknowledgements}


%
%
\bibliographystyle{aa}
\bibliography{paper.bib} 

\end{document}